\pdfoutput=1

\documentclass[apl,amsmath,reprint,floatfix,a4paper,superscriptaddress]{revtex4-1}

\usepackage{graphicx}
\usepackage{dcolumn}
\usepackage{bm}
\usepackage[utf8]{inputenc}
\usepackage[T1]{fontenc}
\usepackage{mathptmx}
\usepackage{textcomp}
\usepackage{booktabs}
\usepackage{upgreek}

\begin{document}

\title{Microscopic origin of magnetization reversal in  exchange-coupled ferro-/ferrimagnetic bilayers} 

\author{Michael Heigl}
\email{michael.heigl@physik.uni-augsburg.de}
\affiliation{Institute of Physics, University of Augsburg, Augsburg 86159, Germany}
\author{Christoph Vogler}
\affiliation{Faculty of Physics, University of Vienna, Boltzmanngasse 5, 1090 Vienna, Austria}
\author{Andrada-Oana Mandru}
\affiliation{Empa--Swiss Federal Laboratories for Materials Science and Technology, CH-8600 Dübendorf, Switzerland}
\author{Xue Zhao}
\affiliation{Empa--Swiss Federal Laboratories for Materials Science and Technology, CH-8600 Dübendorf, Switzerland}
\author{Hans Josef Hug}
\affiliation{Empa--Swiss Federal Laboratories for Materials Science and Technology, CH-8600 Dübendorf, Switzerland}
\affiliation{Department of Physics, University of Basel, CH-4056 Basel, Switzerland}
\author{Dieter Suess}
\affiliation{Christian Doppler Laboratory for Advanced Magnetic Sensing and Materials, Faculty of Physics, University of Vienna, Boltzmanngasse 5, 1090 Vienna, Austria}
\author{Manfred Albrecht}
\affiliation{Institute of Physics, University of Augsburg, Augsburg 86159, Germany}

\begin{abstract}
In this study, the magnetic reversal process of exchange-coupled bilayer systems, consisting of a ferrimagnetic TbFeCo alloy layer and a ferromagnetic [Co/Ni/Pt]$_N$ multilayer, was investigated. In particular, minor loop studies, probing solely the reversal characteristics of the softer ferromagnetic layer, reveal two distinct reversal mechanisms, which depend strongly on the thickness of the ferromagnetic layer. For thick layers, irreversible switching of the macroscopic minor loop is observed. The underlying microscopic origin of this reversal process was studied in detail by high-resolution magnetic force microscopy, showing that the reversal is triggered by in-plane domain walls propagating through the ferromagnetic layer. In contrast, thin ferromagnetic layers show a hysteresis-free reversal, which is nucleation-dominated due to grain-to-grain variations in magnetic anisotropy of the Co/Ni/Pt multilayer and an inhomogeneous exchange coupling with the magnetically hard TbFeCo layer, as confirmed by micromagnetic simulations.
\end{abstract}

\maketitle 


\section{Introduction}

The concept of engineering exchange-coupled composites is the most promising approach to meet current challenges in fabricating high energy density permanent magnets \cite{Fischbacher2017,Zeng_2019, KIM2019139}. Already in 1991, Kneller and Hawig \cite{102931} proposed to manufacture magnets with a magnetically hard and soft phase, exchange-coupled at a mutual interface. While the high magnetocrystalline anisotropy of the hard phase provides a high coercive field, the coupled soft phase should contribute to the energy density product by a high saturation magnetization. Due to the soft phase, the demagnetization curve shows a completely reversible part, which led to the term ‘exchange-spring magnets’. Furthermore, exchange-coupled systems employing ferrimagnetic (FI) heavy rare earth (RE) - 3$d$ transition metal (TM) alloys provide high tunability, interfacial exchange interaction, and zero magnetic moment at the compensation temperature $T_\mathrm{comp}$ \cite{Ungureanu2010,Hauet2007,Patra2009,S.RomerM.A.MarioniK.ThorwarthN.R.JoshiC.E.CorticelliH.J.HugS.OezerM.Parlinska-Wojtan2017,Tokunaga1990,Hauet2009, watson_interfacial_2008,Mangin2003,Mangin2006,Mangin2008,lin_high_2003,Canet2000,PhysRevB.93.184423}, which is highly beneficial for many applications such as spin valves \cite{Dieny1991,Radu2012,doi:10.1063/1.366995,doi:10.1063/1.1851954,doi:10.1063/1.3536476,doi:10.1063/1.2172193} and magnetic tunnel junction devices \cite{Moodera1995}. Below $T_\mathrm{comp}$, the magnetic moment of the RE atoms dominates, which leads to an antiparallel alignment of the net magnetic moments of the FI layer when coupled to a ferromagnetic (FM) layer. As a consequence, a positive horizontal shift of the hysteresis loop of the magnetically softer layer is typically observed after saturation in the positive field direction. In this configuration, a giant exchange bias shift of several Tesla has been reported for various FI/FM bilayer systems which can differ in the reversal behavior of the soft layer exhibiting either fully reversible or irreversible switching \cite{watson_interfacial_2008,romer_temperature_2012,schubert_interfacial_2013,Hebler2017,Vaskovskiy2012}. \\
Micromagnetic simulations showed that a partial domain wall is formed at the FI/FM interface layer during the reversal of the FM \cite{vogler2020hysteresisfree}. The minor loop becomes fully reversible if this domain wall generates a hard-axis field that  overcomes the anisotropy field of the FM. It was further reported that the anisotropy and the bulk exchange of both layers, as well as the exchange coupling strength and the thickness of the FM play an important role in the reversibility of the FM. The underlying reversal mechanism for reversible switching was studied in detail in a [Co(0.4)/Pt(0.7)]$_5$ multilayer exchange-coupled to ferrimagnetic Tb$_{26.5}$Fe$_{73.5}$, that acts as a magnetically hard pinning layer. There, a nucleation-dominated magnetization reversal process was revealed which is caused by grain-to-grain variations in magnetic anisotropy of the Co/Pt multilayer (ML) and an inhomogeneous exchange coupling to the magnetically hard TbFe layer \cite{zhao_magnetization_2019}.\\
In a recent study on a TbFeCo/[Co/Ni/Pt]$_N$ sample series, we have systematically investigated the reversal behavior of the softer FM layer as a function of its thickness, including a detailed theoretical analysis of the full reversibility condition \cite{vogler2020hysteresisfree}. In the present work, we study in detail two specific FI/FM samples out of this sample series exhibiting two distinct cases of reversible and irreversible switching. The different underlying microscopic reversal processes are investigated using high-resolution magnetic force microscopy and micromagnetic simulations. 

\section{Experimental details}

Film deposition was performed at room temperature by dc magnetron sputtering from elemental targets on a Si(001) substrate with a 100\,nm thick thermally oxidized SiO$_x$ layer. The sputter process was carried out using an Ar working pressure of 5x10$^{-3}$\,mbar in an ultra-high vacuum chamber (base pressure < $10^{-8}$\,mbar). The heterostructures consist of a 20\,nm-thick amorphous ferrimagnetic Tb$_{28}$Fe$_{58}$Co$_{14}$ layer and a ferromagnetic [Co(0.2\,nm)/Ni(0.4\,nm)/Pt(0.6\,nm)]$_N$ ML on top. In addition,  5\,nm-thick Pt seed and cover layers were used. The thicknesses of the layers were estimated from the areal densities measured by a quartz balance during deposition while the elemental composition of the TbFeCo alloy was evaluated by Rutherford backscattering spectrometry. Two FI/FM heterostructures with different repetition number ($N$ = 5 and 9) of the ferromagnetic multilayer were chosen for this study. Furthermore, reference samples of the individual layers were prepared.            
\\
The integral magnetic properties of the FI/FM heterostructures were investigated by superconducting quantum interference device - vibrating sample magnetometry (SQUID-VSM). All $M-H$ minor and full loops were measured at 40\,K in the out-of-plane geometry revealing for both systems strong perpendicular magnetic anisotropy where the ferrimagnetic layer acts as magnetically hard pinning layer. Furthermore, the ferrimagnet is Tb dominant over the entire temperature range that we investigated, meaning that the Tb magnetic moment is always larger than the total Co/Ni moment. Consequently, the net moments of the FI/FM heterostructures are antiferromagnetically aligned in the ground state at zero field. The magnetic properties of the full sample series can be found in \cite{vogler2020hysteresisfree}. More details on the magnetic properties of Co/Ni/Pt MLs can be found in \cite{doi:10.1063/5.0010112}.\\
Complementary, the complex reversal behavior was locally imaged at 40\,K by an ultra-high vacuum magnetic force microscope (MFM) operating in magnetic fields of up to 7\,T \cite{doi:10.1063/1.1149970}. Details on the MFM data acquisition and data processing can be found in \cite{Zhao:2018kd, zhao_magnetization_2019}.

\section{Experimental results}

\begin{figure}
\includegraphics[width=8.5cm]{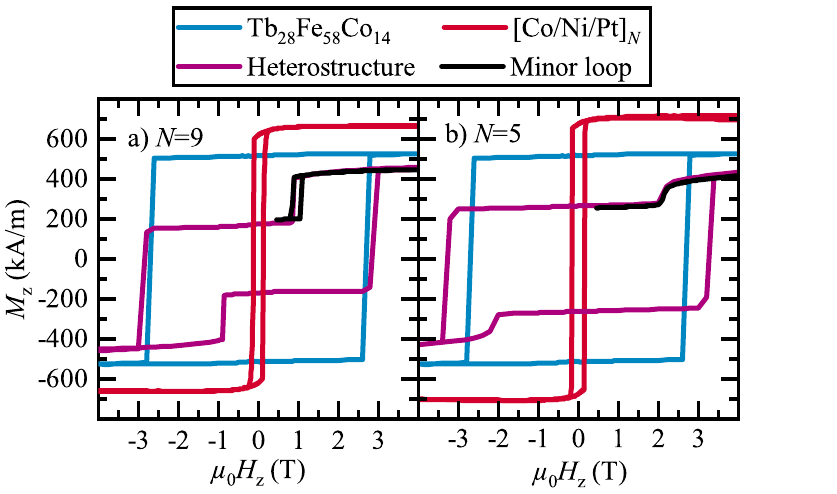}
\caption{\small $M-H$ hysteresis loops of the FI/FM heterostructures obtained at 40\,K. In a) the thicker FM layer ($N$ = 9) switches irreversibly as indicated by the presence of a hysteresis in the minor loop (black). In contrast, the thinner FM layer ($N$ = 5) exhibits a fully reversible switching as shown by the  minor loop (black) in b). The $M-H$ hysteresis loops of the individual layers forming the heterostructures are displayed as well.}
\label{fig:Loops}
\end{figure}

\begin{figure*}
\includegraphics[width=17cm]{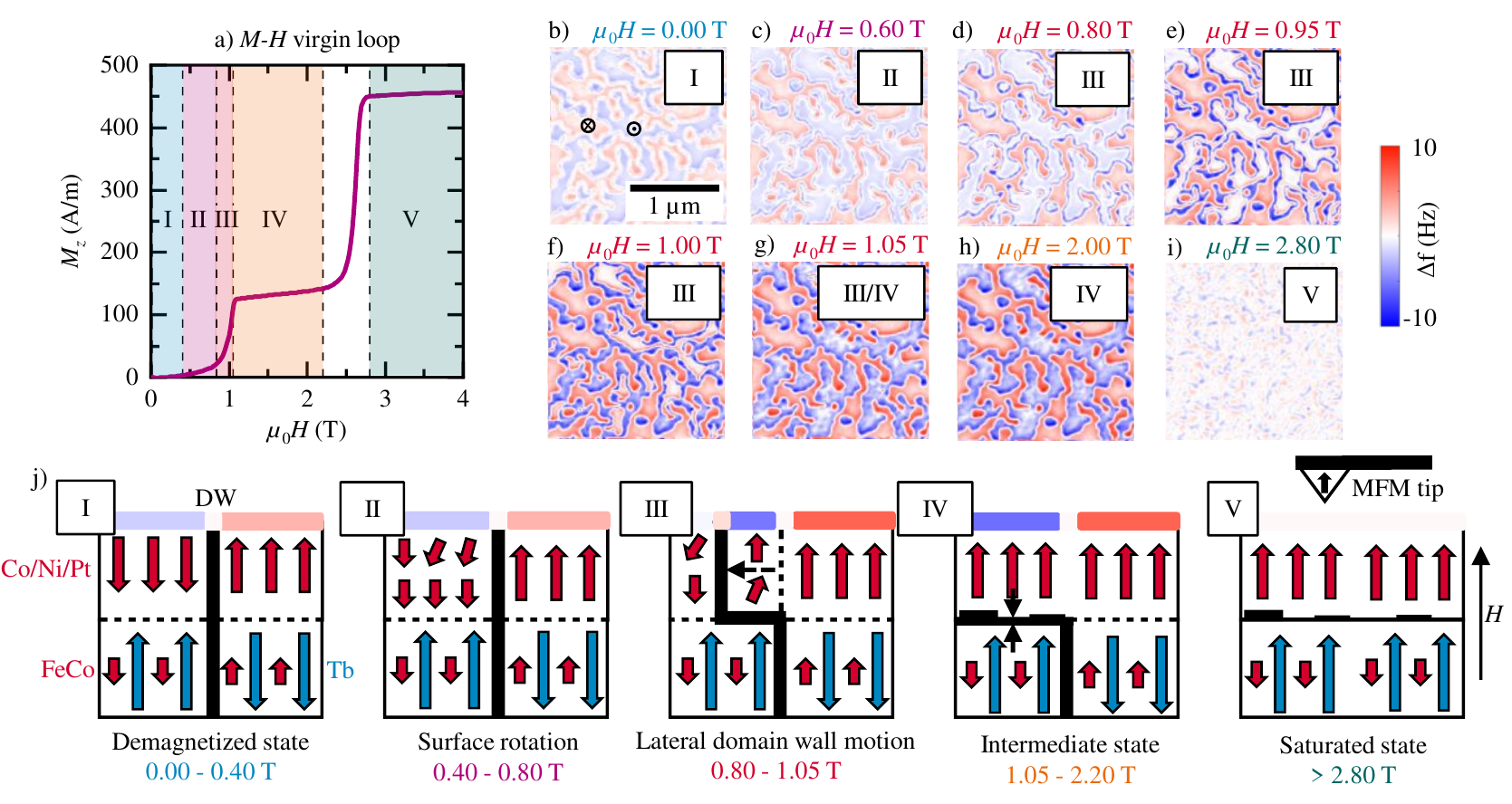}
\caption{\small a) Virgin $M-H$ curve of the FI/FM heterostructure with $N$ = 9 taken at 40\,K starting from the demagnetized state. Five different magnetic states are marked with color. In b-i) in-field MFM images (2\,\textmu m x 2\,\textmu m) are displayed with an applied magnetic field ranging from zero up to 2.80\,T. The five magnetic states, corresponding to the MFM images, are illustrated in j).}
\label{fig:MFM}

\includegraphics[width=17cm]{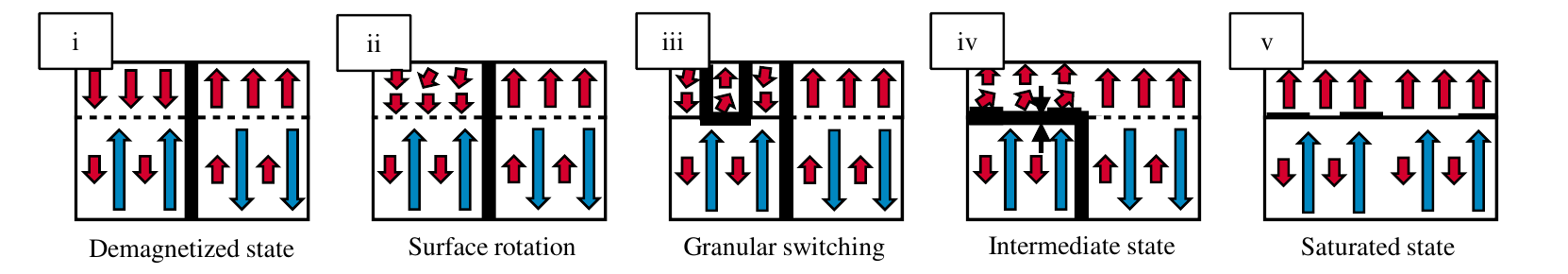}
\caption{\small Schematic showing the underlying microscopic reversal mechanism for fully reversible switching via granular nucleation \cite{zhao_magnetization_2019}.}
\label{fig:Process}
\end{figure*}

The $M-H$ full hysteresis and also minor loops of the exchange-coupled FI/FM systems and of reference samples consisting only of the FI or FM layers recorded at 40\,K are shown in Fig.\,\ref{fig:Loops}. Both reference samples show strong perpendicular magnetic anisotropy with coercive fields of about 3\,T for the FI layer and 200\,mT for the Co/Ni/Pt MLs. 
For the exchanged-coupled FI/FM heterostructures, starting from saturation, by lowering the magnetic field the FM layer reverses due to the strong antiferromagnetic coupling. At a high opposite  field of about 3 T eventually the magnetically hard FI layer switches. It is observed that the field required for reversing the FM layer becomes larger with decreasing number $N$ (thickness), which is expected provided that the interfacial exchange-coupling remains constant \cite{vogler2020hysteresisfree, doi:10.1063/1.3309417}. The minor loops were captured to analyze the switching process of the softer FM layer and show two distinct switching mechanisms. While the FI/FM heterostructure with the thick FM layer exhibits a hysteretic reversal process (Fig.\,\ref{fig:Loops}\,a), the thinner FM layer reveals fully reversible switching (Fig.\,\ref{fig:Loops}\,b). 
\\
In order to get an understanding of the underlying hysteretic irreversible process and to extract differences to the non-hysteretic case, the hysteretic magnetization process occurring for the thicker ($N=9$) FM layer is discussed. For this, the sample was demagnetized to acquire the virgin $M-H$-dependence at 40\,K for fields raised from 0 to 4\,T (Fig.\,\ref{fig:MFM}\,a). A different piece of the same sample in its as-grown state was used to acquire a series of MFM images at 40\,K in fields from 0 to 2.8\,T (Fig.\,\ref{fig:MFM}\,b-i). The magnetic domain structure in the as-grown state is presented in Fig.\,\ref{fig:MFM}\,b. In this state, the net magnetic moment of the FI layer and the magnetic moment of the FM layer are aligned antiparallel, which is the ground state of this exchange coupled FI/FM system. Because the tip magnetization is along the positive field direction (up), the domains appearing with a negative frequency shift contrast (blue domains) have a net up magnetization (parallel to that of the tip), while the red domains have a net down magnetization. The domains are separated by vertical domain walls going through both layers, as schematically displayed in Fig.\,\ref{fig:MFM}\,j,\,I. The magnetization loops acquired for the FI/FM heterostructure sample (purple loop in Fig.\,\ref{fig:Loops}\,a and virgin loop displayed in Fig.\,\ref{fig:MFM}\,a reveal that the Tb magnetic moment dominates the Co and Fe moments of the FI layer and also those of the FM layer. Hence, the blue domains are locations characterized by an up magnetization of the FI layer and a down magnetization of the FM layer. When the field is increased to 0.6\,T (Fig.\,\ref{fig:MFM}\,c) a slight increase of the MFM contrast is observed which is compatible with a small rotation of the magnetic moments at the top of the FM layer away from the initial down direction towards the up direction schematically shown in Fig.\,\ref{fig:MFM}\,j,\,II. A more noticeable change becomes apparent at 0.8\,T (Fig.\,\ref{fig:MFM}\,d): here some dark blue spots occur at the domain walls and the overall contrast is further increased. The latter is compatible with an increase of the rotation of the magnetic moments at the top of the FM layer. The dark blue spots near the domain walls indicate the beginning of a lateral domain wall motion (Fig.\,\ref{fig:MFM}\,j,\,III) that becomes clearly visible in Figs.\,\ref{fig:MFM}\,e and \ref{fig:MFM}\,f for a field of 0.95 and 1\,T, respectively. At about 1.05\,T (Fig.\,\ref{fig:MFM}\,g), the FM layer is almost fully reversed. The remaining MFM contrast then predominantly arises from the domains of the FI layer. For fields between 1.05 and 2\,T, the MFM contrast increases only very slightly, which is compatible with a vertical compression of the horizontal domain wall that has formed at the locations of the blue domains at the FI/FM interface (Fig.\,\ref{fig:MFM}\,j,\,IV). Note however that no further lateral domain wall motion occurs. This reflects the high coercivity of the FI layer. The last MFM image (Fig.\,\ref{fig:MFM}\,i) shows a state where all down domains in the FM have been erased and an in-plane interfacial domain wall has formed (Fig.\,\ref{fig:MFM}\,j,\,V) \cite{schubert_interfacial_2013}. The weak granular contrast is caused by variations of the z-component of the magnetic moment density arising for example from spatial film inhomogeneities (e.g., in the TbFeCo composition \cite{10.3389/fmats.2016.00008, d1a6f4445d5142d39656b29e436b4366}).
\\
The details of the underlying reversal mechanism of the fully reversible switching case were already reported for a similar TbFe/[Co/Pt]$_5$ heterostructure \cite{zhao_magnetization_2019}. There, a nucleation-dominated three-stage magnetization reversal process was revealed which is caused by grain-to-grain variations in magnetic anisotropy of the Co/Pt ML and an inhomogeneous exchange coupling to the magnetically hard TbFe layer. The reversal steps are schematically illustrated in Fig.\,\ref{fig:Process}. They consist of a rotation-dominated part of the FM layer starting at the top surface of individual grains (ii) followed by the full reversal (iii) till saturation (iv). The last reversal step is again characterized by a rotation part of the FI domains till saturation (v).

\section{Micromagnetic Simulations}
In this section, finite-element simulation results obtained with the finite-elements package \textit{magnum.fe}\,\cite{abert_magnum.fe_2013} are presented. This is to reproduce the different minor loop behaviors observed for thin and thick FM layers and from that gain a more fundamental understanding of the relevant physics governing the experimental observations. 
\begin{figure}
\includegraphics[width=8.5cm]{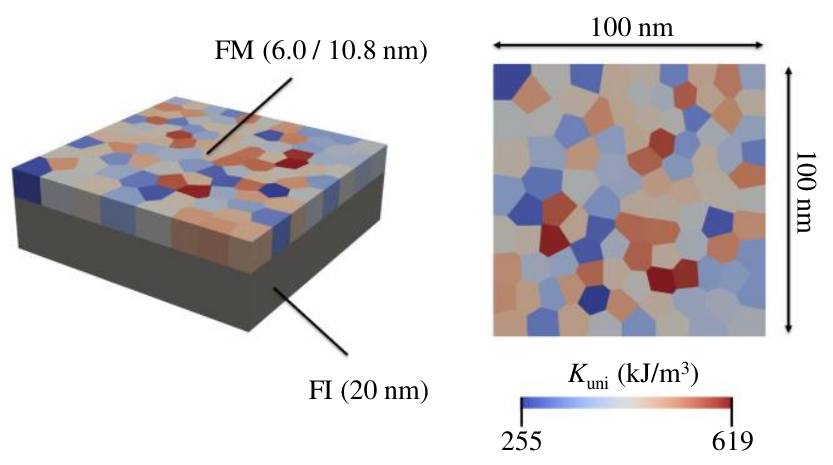}
\caption{\small Model of the simulated FI/FM bilayer structure. The lateral dimension of the model and the thickness of the layers are given. The average grain diameter is 10\,nm and the discretization length is 2\,nm. The color code shows the effective magnetic anisotropy which is assumed to be normally distributed over the grains in the FM layer.}
\label{fig:model_film}
\end{figure}
As shown in Fig.\,\ref{fig:model_film}, we model $100\,{\rm nm}\times 100\,$nm films consisting of a 20\,nm-thick FI layer and FM layers of 6.0 and 10.8\,nm thickness, corresponding to the two different repetition numbers of $N = 5$ and $N=9$, respectively. The structures consist of grains having an average diameter of 10 nm, produced by Voronoi tessellation. Due to the required small discretization length of 2\,nm and the resulting large computational effort, the lateral dimensions are kept at $100\,{\rm nm}\times 100\,$nm which is considerably smaller than the typical domain structures observed by MFM, displayed in Fig.\,\ref{fig:MFM}b-i. However, given that single grains with a lateral length of 50\,nm were well approximated with a spin chain model in a previous work\,\cite{zhao_magnetization_2019}, a homogeneous magnetization within this distance in the in-plane direction can be assumed. Hence, a scaling of the lateral film dimensions is justified. The grains of our model are fully exchange coupled in the lateral directions. The uniaxial magnetic anisotropy $K_u$ is normally distributed with a mean of 430\,kJ/m$^3$ and a standard deviation of 75\,kJ/m$^3$ (total range of 255 to 619\,kJ/m$^3$), and the ratio $J_{\mathrm{iex}}/K_{\mathrm{eff,FM}}$ of interface exchange constant between the two layers and the respective effective magnetic anisotropy constant of the FM is kept constant at -0.1\,$\upmu$m (see color code in Fig.\,\ref{fig:model_film}). All other material parameters are given in Table\,\ref{tab:mat}.
\begin{table}
  \centering
  \vspace{0.5cm}
  \begin{tabular}{c c c}
    \toprule
    \toprule
     Layer & FI & FM \\
    \midrule
    $K_{\mathrm{u}}$\,(kJ/m$^3$) & 1168 & $255-619$ \\
    $M_{\mathrm{S}}$\,(kA/m) & 517 & 668 \\
    $A_{\mathrm{ex}}$\,(pJ/m) & 10.0 & 10.0 \\
    $J_{\mathrm{iex}}$\,(mJ/m$^2$) & \multicolumn{2}{c}{-0.1 to -33.8} \\
    $\alpha$ & 1.0 & 1.0 \\
    $\angle_{\boldsymbol{K}_{\mathrm{eff}},\boldsymbol{e}_z}$\,($^\circ$) & 1.0 & 1.0\\
    \midrule
    $a$\,(nm) & 100.0 & 100.0 \\
    $t$\,(nm) & 20.0 & 6.0 / 10.8\\
    \bottomrule
    \bottomrule
  \end{tabular}
  \caption{\small Material parameters used for the micromagnetic simulations of the investigated 
  FI/FM heterostructures. $K_{\mathrm{u}}$ is the uniaxial magnetic anisotropy constant, $M_{\mathrm{S}}$ is the saturation magnetization, $A_{\mathrm{ex}}$ is the exchange coupling in the bulk, $J_{\mathrm{iex}}$ is the interface exchange coupling between the antiferromagnetically coupled layers, $\alpha$ is the damping constant, $a$ is the lateral size of the geometry and $t$ is the thickness of the layers. The anisotropy axis is tilted by 1\,$^\circ$ against the z-direction in both layers to avoid metastable states.}
\label{tab:mat}
\end{table}

The assumptions of a strong variation in the magnetic anisotropy of the FM layer and simultaneously in the interface-exchange constant at the FI/FM interface are based on previous results reported in Refs.\,\cite{zhao_magnetization_2019,vogler2020hysteresisfree}. In Ref.\,\cite{zhao_magnetization_2019}, a hysteresis-free minor loop was found for a similar heterostructure with a thin FM layer of 5.5\,nm and in Ref.\,\cite{vogler2020hysteresisfree}, the 
relevant parameters and a condition for hysteresis-free minor loops were derived.

For the modeling, the ground state of the heterostructure with the FI layer magnetization pointing in the +z direction and that of the FM layer pointing in the -z direction is considered. Subsequently, the field magnitude is increased step-wise in $50$\,mT increments up to 4\,T and then decreased back to 0\,T. After each field-step the micromagnetic state of the system is relaxed for 1\,ns. Note that the variation of the applied field in the simulations is performed much faster than that used during the experiments. However, because of the high damping constant ($\alpha=1.0$) used for the simulations, a stationary state is obtained within 1\,ns, such that the modeled loops are representative of the experimental loops.

\begin{figure}
\includegraphics[width=8.5cm]{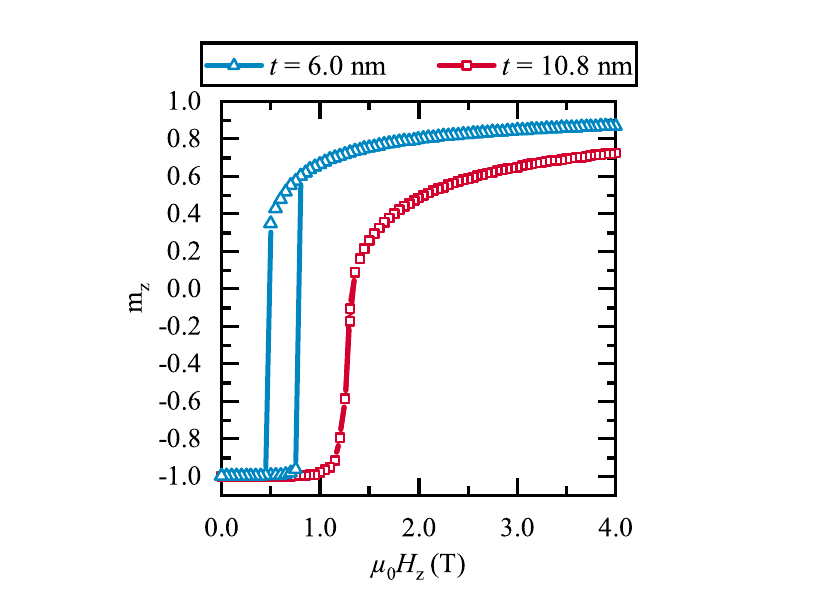}
\caption{\small Simulated macroscopic minor loops of the FM layer coupled to the FI for two different FM thicknesses of 6\,nm ($N=5$, blue line and triangles) and 10.8\,nm ($N=9$, red lines and squares).}
\label{fig:hyst}
\end{figure}
Figure\,\ref{fig:hyst} displays minor loops of heterostructures with a thick ($N=9$) and thin ($N=5$) FM layer. While the loop for the thick FM layer shows a lower switching field of about 580\,mT and a hysteresis width around 300\,mT, the switching field for the thin FM layer is higher than 1\,T and hysteresis is absent. This reproduces the experimentally observed switching behavior of the FM layer (see black lines in Figs.\,\ref{fig:Loops}\,a and b). Note that for the modeling the only parameter changed was the thickness of the FM layer.

\begin{figure*}
    \centering
    \includegraphics{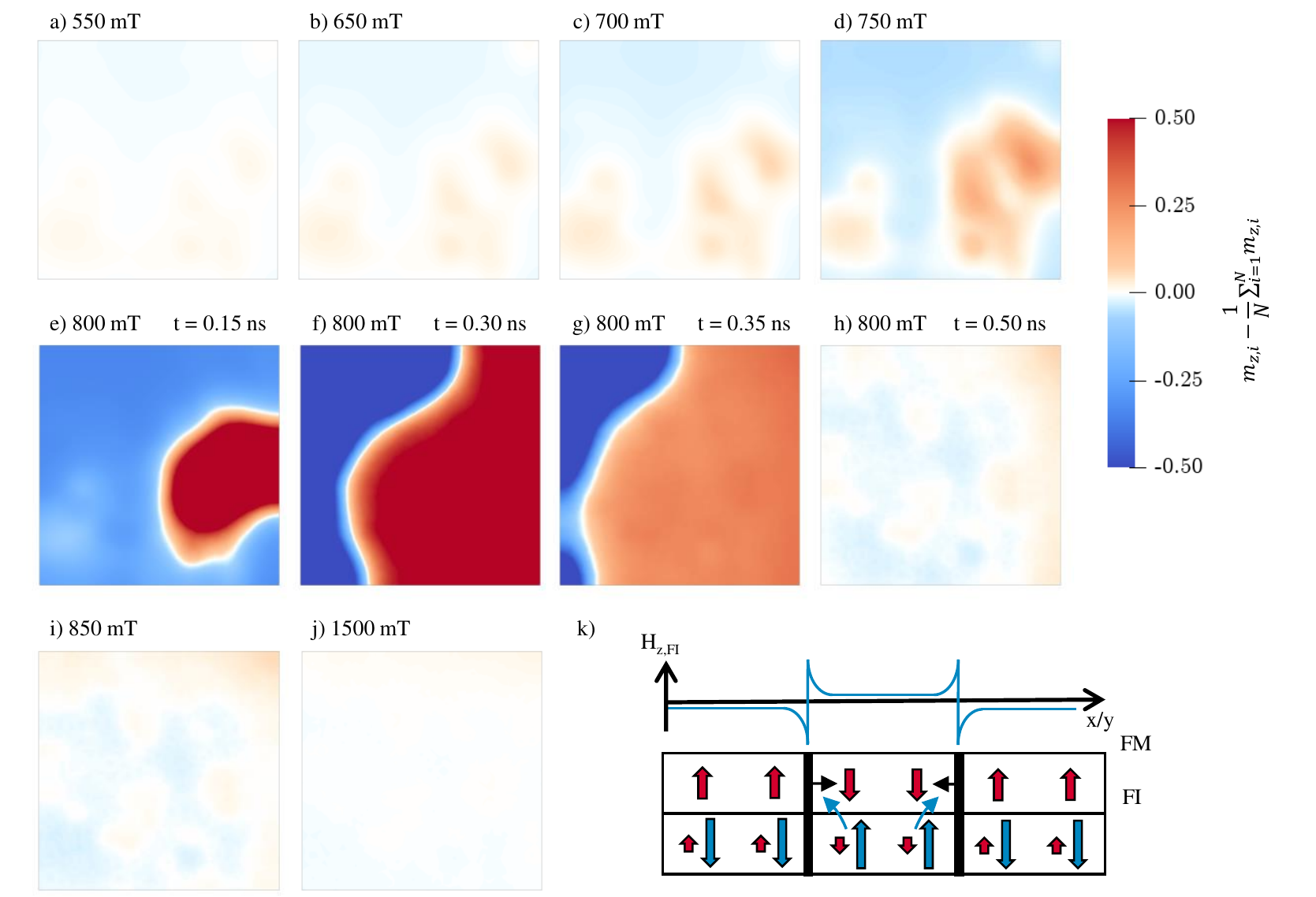}
\caption{\small Simulated MFM contrast of the thicker FM layer of 10.8\,nm ($N=9$) during its reversal for representative chosen external fields. The color code represents the quantity in Eq\,\ref{eq:quant}. e-h) show the dynamic process of the domain wall motion during switching at 800\,mT and not the relaxed magnetization state, as indicated by the simulation time. k) schematically shows the additional local field $H_\mathrm{z,FI}$ arising from the FI layer which ends the lateral domain wall propagation in the FM (Fig.\,\ref{fig:MFM}\,d-g), in contrast to the simulation in e-h).}
\label{fig:MFMsim_thick}
\end{figure*}

To gain a more fundamental understanding of the reversal process of the FM layer and to understand the MFM contrast evolution with the field (Fig.\,\ref{fig:MFM}b-i), the quantity
\begin{equation}
\label{eq:quant}
  m_{z,i}-\frac{1}{N}\sum_{i=1}^Nm_{z,i},
\end{equation}
is plotted, with $m_{z,i}$ being the z-component of the normalized magnetization of node $i$ of the finite-element discretization and with the sum running over all N nodes of the surface mesh on the top of the FM. Note that because a homogeneous magnetic moment distribution does not generate a stray field and hence has no MFM contrast, the average $m_z$ component is subtracted in equation\,\ref{eq:quant} to facilitate the comparison of the simulation results with MFM data.  Figure\,\ref{fig:MFMsim_thick} displays the computed field evolution of the contrast for the sample with thicker FM layer ($N=9$, 10.8\,nm-thick). At low positive fields the top FM layer couples antiparallel to the bottom FI. Therefore, in Fig.\,\ref{fig:MFMsim_thick}\,a-d a weak contrast appears, which can be attributed to a difference in the canting of the magnetic moments in the individual grains with different magnetic anisotropy. The moments in grains with small anisotropy show more canting than those in grains with higher anisotropy. The contrast further increases with increasing external field. Note that the simulation is performed only for the domain with an up magnetization of the FI layer and a down magnetization of the FM layer, because the opposite domain, with an up FM magnetization would not be affected by an applied field well below the coercivity of the FI layer. The simulation thus reveals a continuous raise of the up magnetization of the FM down domain and hence a decrease of the magnetization difference between the FM down and up domains. Therefore, the field arising from the pattern of domains in the FM that is antiparallel to that of the FI layer is reduced, and consequently the MFM contrast is increased. This explains the small increase in contrast observed when comparing the MFM images from Fig.\,\ref{fig:MFM}\,b and c.

At a field of about 800\,mT, a domain wall forms throughout the whole film, starting from the grain with the lowest anisotropy, as illustrated in Fig.\,\ref{fig:MFMsim_thick}\,e-h. The domain wall propagates through the film until all parts have switched parallel to the field (and parallel to the FI net magnetization), which is a fully irreversible process in agreement with the experimental results. Note that since in the relaxed magnetization state at 800\,mT the domain wall propagation cannot be seen, Fig.\,\ref{fig:MFMsim_thick}\,e-h) shows the dynamic process at 800\,mT, as indicated by the simulation time in the upper right corner. In order to compare the simulation results with the observed MFM data some limitations of the simulation work need to be further elaborated. To keep the computational effort at an acceptable level, the simulation considers only a small area within an initially down FM magnetization. Thus, the magnetostatic energy arising from the up/down FM and down/up FI domain pattern and the existence of an initial vertical domain wall inside the FI and FM is not considered. For this reason the simulation reveals the switching of a low anisotropy grain followed by a rapid expansion of the reversal domain. In contrast, the MFM data recorded in fields from 0.8 to 1.05\,T (Fig.\,\ref{fig:MFM}\,d-g) show steady-states of the domain wall propagation that cannot be observed in the model used for simulation. The MFM images reveal that no domain reversal occurs inside the FM down domain, but that the reversal starts by the propagation of an already existing wall and that the propagating wall again becomes pinned for fields from 0.8\,T to 1.05\,T. This can be explained by the up field arising from the FI up domains, that is strongest in the inside of the FM down domain near its wall (Fig.\,\ref{fig:MFMsim_thick}\,k). This stray field from the FI layer adds to the applied up external field and thus drives the propagation of the existing wall to a location inside the FI up domain where the up stray field from the FI domain pattern is weaker. Because our simulation considers a FM down domain only, this behavior cannot be modeled. The simulation however reveals that once a reversal domain exists, a rapid wall propagation follows. This explains why the field-interval where a domain wall propagation is observed remains small, i.e. about 0.2\,T which is compatible with the field on the surface of the FI layer near a domain wall. 

After the FM has switched, the contrast becomes abruptly weak and decreases with increasing external field, as displayed in Fig.\,\ref{fig:MFMsim_thick}\,i-j. Note that again only the FM magnetization inside an initial FM down domain is considered. The simulation shows that the grain-to-grain variation of the up magnetic moment is reduced with increasing up field, which corresponds to a compression of the in-plane domain wall that has formed at the FI/FM interface. Hence, the up magnetic moment of the FM layer is increased approaching that of a FM domain with an initial up magnetization. The down/up field from the FM layer that weakened the up/down field of the domains inside the FI layer thus becomes gradually smaller. This explains the small increase of the MFM contrast observed when the field is increased from 1.05 to 2\,T (Figs.\,\ref{fig:MFM}\,g and \ref{fig:MFM}\,h).

\begin{figure*}
    \centering
    \includegraphics[width=17cm]{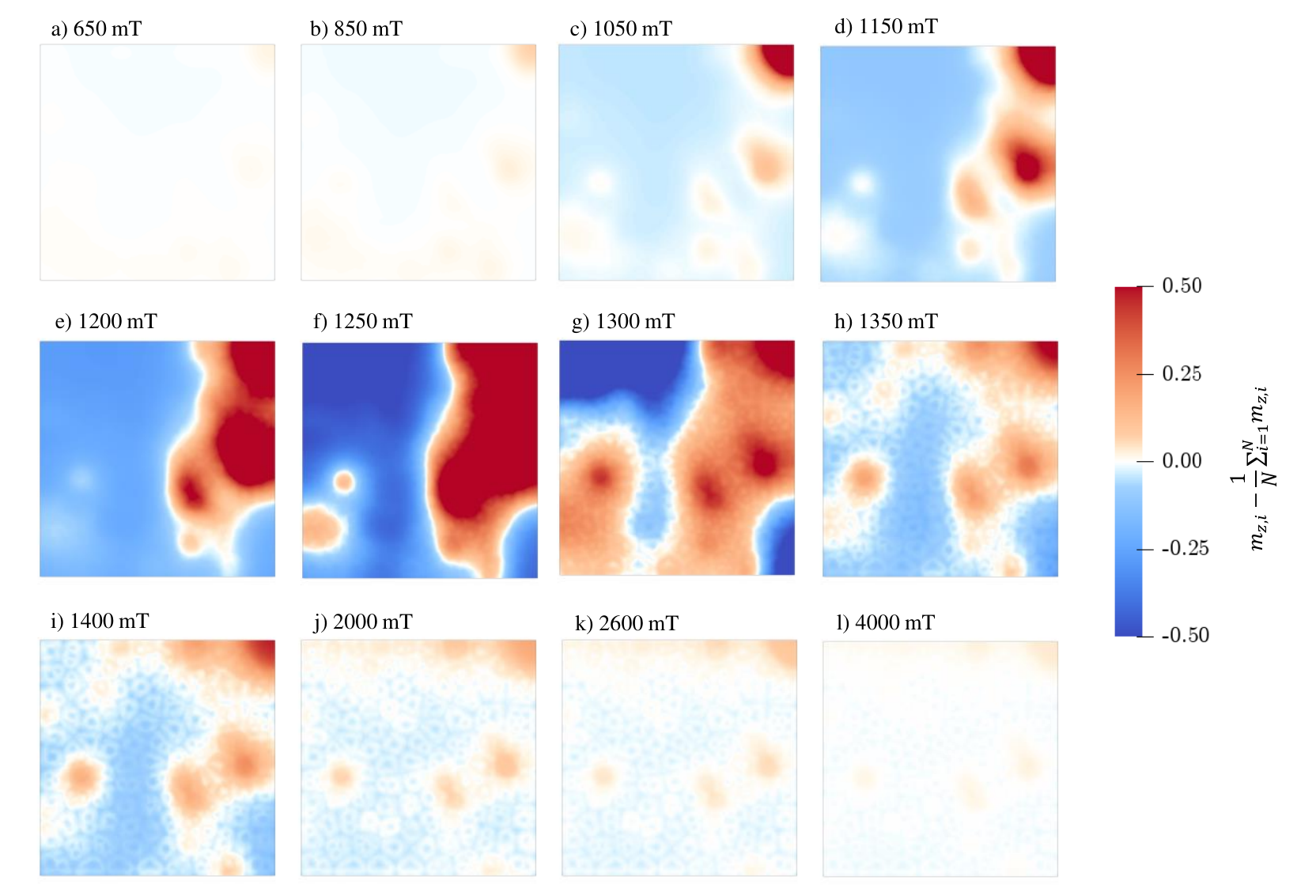}
\caption{\small Simulated MFM contrast of the thinner FM layer of 6\,nm ($N=5$) during its reversal for representative chosen external fields. The color code represents the quantity of Eq\,\ref{eq:quant}.}
\label{fig:MFMsim_thin}
\end{figure*}

Note that the magnetization process observed here for the thicker FM layer is fundamentally different from that observed in our previous work\,\cite{zhao_magnetization_2019} for a thinner FM layer. For this reason, the modeling is also performed for the thinner FM layer using the parameters describing the FI/FM heterostructure samples fabricated here. The results are displayed in Fig.\,\ref{fig:MFMsim_thin}. Again the modeling is performed for a film area with an initial down FM magnetization. As in our previous work\,\cite{zhao_magnetization_2019} a 3-stage magnetization process is observed. Stage 1 (panels a-c) is characterized by a rotation of the initially down magnetic moments near the top of the FM layer towards the up direction of the applied field. In stage 2 (panels d-g), the magnetic moments of isolated grains switch towards the up direction to improve the alignment of the magnetic moments to the applied field. The angle between the magnetic moments and the field is then smaller for the moments near the top of the FM film and larger for the moments near the FI/FM interface. Because this switching process depends on the properties of individual grains, a large grain-to-grain variation occurs. In stage 3 (panels h-l) all grains have switched to have a predominately up magnetization. The contrast drops as the horizontal domain wall at the FI/FM interface is compressed and the local variation of the domain wall thickness is decreased. As observed by MFM in our former work\,\cite{zhao_magnetization_2019}, no lateral propagation of a vertical domain wall inside the FM layer occurs, but a three-stage magnetization process takes place for each individual grain. The reversal process is hysteresis free (red curve with squares in Fig.\,\ref{fig:hyst}).

The condition for hysteresis-free switching derived in Ref.\,\cite{vogler2020hysteresisfree} can be applied here. Using the parameters listed in Table\,\ref{tab:mat} for the thicker FM layer clearly reveals that 90\,\% of the grains in the heterostructure remain above the threshold for hysteresis-free switching ($K_{\mathrm{uni}}\le330$\,kJ/m$^3$), while for the system with the thinner FM the threshold value is $K_{\mathrm{uni}}\le418$\,kJ/m$^3$. This means that 40\/\% of the grains are below the threshold and hence show a hysteresis-free process. For the thinner FI/FM heterostructure with $N=5$ this means that almost half of the FM grains show a gradual, hysteresis-free rotation of the magnetization. Based on this behavior and the small thickness of the FM, short vertical domain walls can form, as schematically illustrated in Fig.\,\ref{fig:Process}, and the observed 3-stage process with a local rearrangement of domains in the switching process occurs. In contrast, in the thick FM layer with $N=9$ almost all grains show irreversible, hysteretic reversal. An abrupt switch of individual grains would generate several large vertical domain walls, which is energetically unfavourable due to the thickness of the FM. This is the reason why we observe only one domain wall, that is independent from the granular structure, propagating through the FM layer and causing irreversible switching of the macroscopic minor loop. 
\\

\section{Summary}
We reveal two distinct magnetic reversal mechanisms in an exchange-coupled bilayer system consisting of a ferrimagnetic TbFeCo alloy layer and a ferromagnetic [Co/Ni/Pt]$_N$ multilayer. The reversal characteristics depend strongly on the thickness of the FM layer. By minor loop $M-H$ measurements, we observed an irreversible hysteretic switching process of the bilayer with $N=9$ and a reversible switching for $N=5$. The underlying microscopic origin is revealed by high-resolution MFM. For $N=9$, the FM switches by in-plane domain wall propagation. In contrast, thinner FM layers exhibit a nucleation-dominated reversal due to grain-to-grain variations in magnetic anisotropy of the Co/Ni/Pt multilayer and an inhomogeneous exchange coupling with the magnetically hard TbFeCo layer. The coupled FM layers of both systems were modeled by finite-element simulations with individual grains varying in $K_\mathrm{u}$. The simulated macroscopic minor loops agreed very well with the experiments. The simulated MFM contrast of the thicker FM layer revealed a dynamic process of the domain wall motion during switching in contrast to the experimentally observed magnetic relaxed states. This difference could be explained by the additional local field arising from the FI layer which is absent in the model. With this exception, the simulations replicated the two switching mechanisms of both systems.

\section*{Acknowledgements}
Financial support for this joint D.A.CH project provided by the German Research Foundation under Grant AL 618/28-1, the Swiss National Science Foundation under Grant 200021-147084, and the Austrian Science Fund under Grant I2214-N20, is gratefully acknowledged. The computational results presented have been achieved using the Vienna Scientific Cluster (VSC).
\\

\section*{\label{sec:Data}Data availability}
The data that support the findings of this study are available from the
corresponding author upon reasonable request.

\bibliography{main}

\begin{thebibliography}{37}%
\makeatletter
\providecommand \@ifxundefined [1]{%
 \@ifx{#1\undefined}
}%
\providecommand \@ifnum [1]{%
 \ifnum #1\expandafter \@firstoftwo
 \else \expandafter \@secondoftwo
 \fi
}%
\providecommand \@ifx [1]{%
 \ifx #1\expandafter \@firstoftwo
 \else \expandafter \@secondoftwo
 \fi
}%
\providecommand \natexlab [1]{#1}%
\providecommand \enquote  [1]{``#1''}%
\providecommand \bibnamefont  [1]{#1}%
\providecommand \bibfnamefont [1]{#1}%
\providecommand \citenamefont [1]{#1}%
\providecommand \href@noop [0]{\@secondoftwo}%
\providecommand \href [0]{\begingroup \@sanitize@url \@href}%
\providecommand \@href[1]{\@@startlink{#1}\@@href}%
\providecommand \@@href[1]{\endgroup#1\@@endlink}%
\providecommand \@sanitize@url [0]{\catcode `\\12\catcode `\$12\catcode
  `\&12\catcode `\#12\catcode `\^12\catcode `\_12\catcode `\%12\relax}%
\providecommand \@@startlink[1]{}%
\providecommand \@@endlink[0]{}%
\providecommand \url  [0]{\begingroup\@sanitize@url \@url }%
\providecommand \@url [1]{\endgroup\@href {#1}{\urlprefix }}%
\providecommand \urlprefix  [0]{URL }%
\providecommand \Eprint [0]{\href }%
\providecommand \doibase [0]{http://dx.doi.org/}%
\providecommand \selectlanguage [0]{\@gobble}%
\providecommand \bibinfo  [0]{\@secondoftwo}%
\providecommand \bibfield  [0]{\@secondoftwo}%
\providecommand \translation [1]{[#1]}%
\providecommand \BibitemOpen [0]{}%
\providecommand \bibitemStop [0]{}%
\providecommand \bibitemNoStop [0]{.\EOS\space}%
\providecommand \EOS [0]{\spacefactor3000\relax}%
\providecommand \BibitemShut  [1]{\csname bibitem#1\endcsname}%
\let\auto@bib@innerbib\@empty
\bibitem [{\citenamefont {Fischbacher}\ \emph {et~al.}(2017)\citenamefont
  {Fischbacher}, \citenamefont {Kovacs}, \citenamefont {Oezelt}, \citenamefont
  {Gusenbauer}, \citenamefont {Schrefl}, \citenamefont {Exl}, \citenamefont
  {Givord}, \citenamefont {Dempsey}, \citenamefont {Zimanyi}, \citenamefont
  {Winklhofer}, \citenamefont {Hrkac}, \citenamefont {Chantrell}, \citenamefont
  {Sakuma}, \citenamefont {Yano}, \citenamefont {Kato}, \citenamefont {Shoji},\
  and\ \citenamefont {Manabe}}]{Fischbacher2017}%
  \BibitemOpen
  \bibfield  {author} {\bibinfo {author} {\bibfnamefont {J.}~\bibnamefont
  {Fischbacher}}, \bibinfo {author} {\bibfnamefont {A.}~\bibnamefont {Kovacs}},
  \bibinfo {author} {\bibfnamefont {H.}~\bibnamefont {Oezelt}}, \bibinfo
  {author} {\bibfnamefont {M.}~\bibnamefont {Gusenbauer}}, \bibinfo {author}
  {\bibfnamefont {T.}~\bibnamefont {Schrefl}}, \bibinfo {author} {\bibfnamefont
  {L.}~\bibnamefont {Exl}}, \bibinfo {author} {\bibfnamefont {D.}~\bibnamefont
  {Givord}}, \bibinfo {author} {\bibfnamefont {N.~M.}\ \bibnamefont {Dempsey}},
  \bibinfo {author} {\bibfnamefont {G.}~\bibnamefont {Zimanyi}}, \bibinfo
  {author} {\bibfnamefont {M.}~\bibnamefont {Winklhofer}}, \bibinfo {author}
  {\bibfnamefont {G.}~\bibnamefont {Hrkac}}, \bibinfo {author} {\bibfnamefont
  {R.}~\bibnamefont {Chantrell}}, \bibinfo {author} {\bibfnamefont
  {N.}~\bibnamefont {Sakuma}}, \bibinfo {author} {\bibfnamefont
  {M.}~\bibnamefont {Yano}}, \bibinfo {author} {\bibfnamefont {A.}~\bibnamefont
  {Kato}}, \bibinfo {author} {\bibfnamefont {T.}~\bibnamefont {Shoji}}, \ and\
  \bibinfo {author} {\bibfnamefont {A.}~\bibnamefont {Manabe}},\ }\href
  {\doibase 10.1063/1.4999315} {\bibfield  {journal} {\bibinfo  {journal}
  {Appl. Phys. Lett.}\ }\textbf {\bibinfo {volume} {111}},\ \bibinfo {pages}
  {072404} (\bibinfo {year} {2017})}\BibitemShut {NoStop}%
\bibitem [{\citenamefont {Zeng}\ \emph {et~al.}(2019)\citenamefont {Zeng},
  \citenamefont {Yu}, \citenamefont {Zhou}, \citenamefont {Zhang},
  \citenamefont {Liao},\ and\ \citenamefont {Liu}}]{Zeng_2019}%
  \BibitemOpen
  \bibfield  {author} {\bibinfo {author} {\bibfnamefont {H.~X.}\ \bibnamefont
  {Zeng}}, \bibinfo {author} {\bibfnamefont {H.~Y.}\ \bibnamefont {Yu}},
  \bibinfo {author} {\bibfnamefont {Q.}~\bibnamefont {Zhou}}, \bibinfo {author}
  {\bibfnamefont {J.~S.}\ \bibnamefont {Zhang}}, \bibinfo {author}
  {\bibfnamefont {X.~F.}\ \bibnamefont {Liao}}, \ and\ \bibinfo {author}
  {\bibfnamefont {Z.~W.}\ \bibnamefont {Liu}},\ }\href {\doibase
  10.1088/2053-1591/ab3756} {\bibfield  {journal} {\bibinfo  {journal} {Mater.
  Res. Express}\ }\textbf {\bibinfo {volume} {6}},\ \bibinfo {pages} {106105}
  (\bibinfo {year} {2019})}\BibitemShut {NoStop}%
\bibitem [{\citenamefont {Kim}\ \emph {et~al.}(2019)\citenamefont {Kim},
  \citenamefont {Sasaki}, \citenamefont {Ohkubo}, \citenamefont {Takada},
  \citenamefont {Kato}, \citenamefont {Kaneko},\ and\ \citenamefont
  {Hono}}]{KIM2019139}%
  \BibitemOpen
  \bibfield  {author} {\bibinfo {author} {\bibfnamefont {T.-H.}\ \bibnamefont
  {Kim}}, \bibinfo {author} {\bibfnamefont {T.}~\bibnamefont {Sasaki}},
  \bibinfo {author} {\bibfnamefont {T.}~\bibnamefont {Ohkubo}}, \bibinfo
  {author} {\bibfnamefont {Y.}~\bibnamefont {Takada}}, \bibinfo {author}
  {\bibfnamefont {A.}~\bibnamefont {Kato}}, \bibinfo {author} {\bibfnamefont
  {Y.}~\bibnamefont {Kaneko}}, \ and\ \bibinfo {author} {\bibfnamefont
  {K.}~\bibnamefont {Hono}},\ }\href {\doibase
  https://doi.org/10.1016/j.actamat.2019.04.032} {\bibfield  {journal}
  {\bibinfo  {journal} {Acta Materialia}\ }\textbf {\bibinfo {volume} {172}},\
  \bibinfo {pages} {139} (\bibinfo {year} {2019})}\BibitemShut {NoStop}%
\bibitem [{\citenamefont {{Kneller}}\ and\ \citenamefont
  {{Hawig}}(1991)}]{102931}%
  \BibitemOpen
  \bibfield  {author} {\bibinfo {author} {\bibfnamefont {E.~F.}\ \bibnamefont
  {{Kneller}}}\ and\ \bibinfo {author} {\bibfnamefont {R.}~\bibnamefont
  {{Hawig}}},\ }\href@noop {} {\bibfield  {journal} {\bibinfo  {journal} {IEEE
  Trans. Magn.}\ }\textbf {\bibinfo {volume} {27}},\ \bibinfo {pages} {3588}
  (\bibinfo {year} {1991})}\BibitemShut {NoStop}%
\bibitem [{\citenamefont {Ungureanu}\ \emph {et~al.}(2010)\citenamefont
  {Ungureanu}, \citenamefont {Dumesnil}, \citenamefont {Dufour}, \citenamefont
  {Gonzalez}, \citenamefont {Wilhelm}, \citenamefont {Smekhova},\ and\
  \citenamefont {Rogalev}}]{Ungureanu2010}%
  \BibitemOpen
  \bibfield  {author} {\bibinfo {author} {\bibfnamefont {M.}~\bibnamefont
  {Ungureanu}}, \bibinfo {author} {\bibfnamefont {K.}~\bibnamefont {Dumesnil}},
  \bibinfo {author} {\bibfnamefont {C.}~\bibnamefont {Dufour}}, \bibinfo
  {author} {\bibfnamefont {N.}~\bibnamefont {Gonzalez}}, \bibinfo {author}
  {\bibfnamefont {F.}~\bibnamefont {Wilhelm}}, \bibinfo {author} {\bibfnamefont
  {A.}~\bibnamefont {Smekhova}}, \ and\ \bibinfo {author} {\bibfnamefont
  {A.}~\bibnamefont {Rogalev}},\ }\href {\doibase 10.1103/PhysRevB.82.174421}
  {\bibfield  {journal} {\bibinfo  {journal} {Phys. Rev. B}\ }\textbf {\bibinfo
  {volume} {82}},\ \bibinfo {pages} {174421} (\bibinfo {year}
  {2010})}\BibitemShut {NoStop}%
\bibitem [{\citenamefont {Hauet}\ \emph {et~al.}(2007)\citenamefont {Hauet},
  \citenamefont {Mangin}, \citenamefont {Montaigne}, \citenamefont {Borchers},\
  and\ \citenamefont {Henry}}]{Hauet2007}%
  \BibitemOpen
  \bibfield  {author} {\bibinfo {author} {\bibfnamefont {T.}~\bibnamefont
  {Hauet}}, \bibinfo {author} {\bibfnamefont {S.}~\bibnamefont {Mangin}},
  \bibinfo {author} {\bibfnamefont {F.}~\bibnamefont {Montaigne}}, \bibinfo
  {author} {\bibfnamefont {J.~A.}\ \bibnamefont {Borchers}}, \ and\ \bibinfo
  {author} {\bibfnamefont {Y.}~\bibnamefont {Henry}},\ }\href {\doibase
  10.1063/1.2753108} {\bibfield  {journal} {\bibinfo  {journal} {Appl. Phys.
  Lett.}\ }\textbf {\bibinfo {volume} {91}},\ \bibinfo {pages} {022505}
  (\bibinfo {year} {2007})}\BibitemShut {NoStop}%
\bibitem [{\citenamefont {Patra}\ \emph {et~al.}(2009)\citenamefont {Patra},
  \citenamefont {Thakur}, \citenamefont {Majumdar},\ and\ \citenamefont
  {Giri}}]{Patra2009}%
  \BibitemOpen
  \bibfield  {author} {\bibinfo {author} {\bibfnamefont {M.}~\bibnamefont
  {Patra}}, \bibinfo {author} {\bibfnamefont {M.}~\bibnamefont {Thakur}},
  \bibinfo {author} {\bibfnamefont {S.}~\bibnamefont {Majumdar}}, \ and\
  \bibinfo {author} {\bibfnamefont {S.}~\bibnamefont {Giri}},\ }\href {\doibase
  10.1088/0953-8984/21/23/236004} {\bibfield  {journal} {\bibinfo  {journal}
  {J. Phys.: Condens. Matter.}\ }\textbf {\bibinfo {volume} {21}},\ \bibinfo
  {pages} {236004} (\bibinfo {year} {2009})}\BibitemShut {NoStop}%
\bibitem [{\citenamefont {{S. Romer, M. A. Marioni, K. Thorwarth, N. R. Joshi,
  C. E. Corticelli, H. J. Hug, S. Oezer, M. Parlinska- Wojtan}}\ and\
  \citenamefont
  {Rohrmann}(2017)}]{S.RomerM.A.MarioniK.ThorwarthN.R.JoshiC.E.CorticelliH.J.HugS.OezerM.Parlinska-Wojtan2017}%
  \BibitemOpen
  \bibfield  {author} {\bibinfo {author} {\bibnamefont {{S. Romer, M. A.
  Marioni, K. Thorwarth, N. R. Joshi, C. E. Corticelli, H. J. Hug, S. Oezer, M.
  Parlinska- Wojtan}}}\ and\ \bibinfo {author} {\bibfnamefont {H.}~\bibnamefont
  {Rohrmann}},\ }\href@noop {} {\bibfield  {journal} {\bibinfo  {journal}
  {Appl. Phys. Lett.}\ }\textbf {\bibinfo {volume} {101}},\ \bibinfo {pages}
  {222404} (\bibinfo {year} {2017})}\BibitemShut {NoStop}%
\bibitem [{\citenamefont {Tokunaga}\ \emph {et~al.}(1990)\citenamefont
  {Tokunaga}, \citenamefont {Taguchi}, \citenamefont {Fukami}, \citenamefont
  {Nakaki},\ and\ \citenamefont {Tsutsumi}}]{Tokunaga1990}%
  \BibitemOpen
  \bibfield  {author} {\bibinfo {author} {\bibfnamefont {T.}~\bibnamefont
  {Tokunaga}}, \bibinfo {author} {\bibfnamefont {M.}~\bibnamefont {Taguchi}},
  \bibinfo {author} {\bibfnamefont {T.}~\bibnamefont {Fukami}}, \bibinfo
  {author} {\bibfnamefont {Y.}~\bibnamefont {Nakaki}}, \ and\ \bibinfo {author}
  {\bibfnamefont {K.}~\bibnamefont {Tsutsumi}},\ }\href {\doibase
  10.1063/1.344917} {\bibfield  {journal} {\bibinfo  {journal} {J. Appl.
  Phys.}\ }\textbf {\bibinfo {volume} {67}},\ \bibinfo {pages} {4417} (\bibinfo
  {year} {1990})}\BibitemShut {NoStop}%
\bibitem [{\citenamefont {Hauet}\ \emph {et~al.}(2009)\citenamefont {Hauet},
  \citenamefont {Montaigne}, \citenamefont {Hehn}, \citenamefont {Henry},\ and\
  \citenamefont {Mangin}}]{Hauet2009}%
  \BibitemOpen
  \bibfield  {author} {\bibinfo {author} {\bibfnamefont {T.}~\bibnamefont
  {Hauet}}, \bibinfo {author} {\bibfnamefont {F.}~\bibnamefont {Montaigne}},
  \bibinfo {author} {\bibfnamefont {M.}~\bibnamefont {Hehn}}, \bibinfo {author}
  {\bibfnamefont {Y.}~\bibnamefont {Henry}}, \ and\ \bibinfo {author}
  {\bibfnamefont {S.}~\bibnamefont {Mangin}},\ }\href {\doibase
  10.1103/PhysRevB.79.224435} {\bibfield  {journal} {\bibinfo  {journal} {Phys.
  Rev. B}\ }\textbf {\bibinfo {volume} {79}},\ \bibinfo {pages} {224435}
  (\bibinfo {year} {2009})}\BibitemShut {NoStop}%
\bibitem [{\citenamefont {Watson}\ \emph {et~al.}(2008)\citenamefont {Watson},
  \citenamefont {Hauet}, \citenamefont {Borchers}, \citenamefont {Mangin},\
  and\ \citenamefont {Fullerton}}]{watson_interfacial_2008}%
  \BibitemOpen
  \bibfield  {author} {\bibinfo {author} {\bibfnamefont {S.~M.}\ \bibnamefont
  {Watson}}, \bibinfo {author} {\bibfnamefont {T.}~\bibnamefont {Hauet}},
  \bibinfo {author} {\bibfnamefont {J.~A.}\ \bibnamefont {Borchers}}, \bibinfo
  {author} {\bibfnamefont {S.}~\bibnamefont {Mangin}}, \ and\ \bibinfo {author}
  {\bibfnamefont {E.~E.}\ \bibnamefont {Fullerton}},\ }\href {\doibase
  10.1063/1.2936836} {\bibfield  {journal} {\bibinfo  {journal} {Appl. Phys.
  Lett.}\ }\textbf {\bibinfo {volume} {92}},\ \bibinfo {pages} {202507}
  (\bibinfo {year} {2008})}\BibitemShut {NoStop}%
\bibitem [{\citenamefont {Mangin}\ \emph {et~al.}(2003)\citenamefont {Mangin},
  \citenamefont {Montaigne},\ and\ \citenamefont {Schuhl}}]{Mangin2003}%
  \BibitemOpen
  \bibfield  {author} {\bibinfo {author} {\bibfnamefont {S.}~\bibnamefont
  {Mangin}}, \bibinfo {author} {\bibfnamefont {F.}~\bibnamefont {Montaigne}}, \
  and\ \bibinfo {author} {\bibfnamefont {A.}~\bibnamefont {Schuhl}},\ }\href
  {\doibase 10.1103/PhysRevB.68.140404} {\bibfield  {journal} {\bibinfo
  {journal} {Phys. Rev. B}\ }\textbf {\bibinfo {volume} {68}},\ \bibinfo
  {pages} {140404(R)} (\bibinfo {year} {2003})}\BibitemShut {NoStop}%
\bibitem [{\citenamefont {Mangin}\ \emph {et~al.}(2006)\citenamefont {Mangin},
  \citenamefont {Hauet}, \citenamefont {Henry}, \citenamefont {Montaigne},\
  and\ \citenamefont {Fullerton}}]{Mangin2006}%
  \BibitemOpen
  \bibfield  {author} {\bibinfo {author} {\bibfnamefont {S.}~\bibnamefont
  {Mangin}}, \bibinfo {author} {\bibfnamefont {T.}~\bibnamefont {Hauet}},
  \bibinfo {author} {\bibfnamefont {Y.}~\bibnamefont {Henry}}, \bibinfo
  {author} {\bibfnamefont {F.}~\bibnamefont {Montaigne}}, \ and\ \bibinfo
  {author} {\bibfnamefont {E.~E.}\ \bibnamefont {Fullerton}},\ }\href {\doibase
  10.1103/PhysRevB.74.024414} {\bibfield  {journal} {\bibinfo  {journal} {Phys.
  Rev. B}\ }\textbf {\bibinfo {volume} {74}},\ \bibinfo {pages} {024414}
  (\bibinfo {year} {2006})}\BibitemShut {NoStop}%
\bibitem [{\citenamefont {Mangin}\ \emph {et~al.}(2008)\citenamefont {Mangin},
  \citenamefont {Hauet}, \citenamefont {Fischer}, \citenamefont {Kim},
  \citenamefont {Kortright}, \citenamefont {Chesnel}, \citenamefont
  {Arenholz},\ and\ \citenamefont {Fullerton}}]{Mangin2008}%
  \BibitemOpen
  \bibfield  {author} {\bibinfo {author} {\bibfnamefont {S.}~\bibnamefont
  {Mangin}}, \bibinfo {author} {\bibfnamefont {T.}~\bibnamefont {Hauet}},
  \bibinfo {author} {\bibfnamefont {P.}~\bibnamefont {Fischer}}, \bibinfo
  {author} {\bibfnamefont {D.~H.}\ \bibnamefont {Kim}}, \bibinfo {author}
  {\bibfnamefont {J.~B.}\ \bibnamefont {Kortright}}, \bibinfo {author}
  {\bibfnamefont {K.}~\bibnamefont {Chesnel}}, \bibinfo {author} {\bibfnamefont
  {E.}~\bibnamefont {Arenholz}}, \ and\ \bibinfo {author} {\bibfnamefont
  {E.~E.}\ \bibnamefont {Fullerton}},\ }\href {\doibase
  10.1103/PhysRevB.78.024424} {\bibfield  {journal} {\bibinfo  {journal} {Phys.
  Rev. B}\ }\textbf {\bibinfo {volume} {78}},\ \bibinfo {pages} {024424}
  (\bibinfo {year} {2008})}\BibitemShut {NoStop}%
\bibitem [{\citenamefont {Lin}\ \emph {et~al.}(2003)\citenamefont {Lin},
  \citenamefont {Lai}, \citenamefont {Jiang},\ and\ \citenamefont
  {Shieh}}]{lin_high_2003}%
  \BibitemOpen
  \bibfield  {author} {\bibinfo {author} {\bibfnamefont {C.-C.}\ \bibnamefont
  {Lin}}, \bibinfo {author} {\bibfnamefont {C.-H.}\ \bibnamefont {Lai}},
  \bibinfo {author} {\bibfnamefont {R.-F.}\ \bibnamefont {Jiang}}, \ and\
  \bibinfo {author} {\bibfnamefont {H.-P.~D.}\ \bibnamefont {Shieh}},\ }\href
  {\doibase 10.1063/1.1556932} {\bibfield  {journal} {\bibinfo  {journal} {J.
  Appl. Phys.}\ }\textbf {\bibinfo {volume} {93}},\ \bibinfo {pages} {6832}
  (\bibinfo {year} {2003})}\BibitemShut {NoStop}%
\bibitem [{\citenamefont {Canet}\ \emph {et~al.}(2000)\citenamefont {Canet},
  \citenamefont {Mangin}, \citenamefont {Bellouard},\ and\ \citenamefont
  {Piecuch}}]{Canet2000}%
  \BibitemOpen
  \bibfield  {author} {\bibinfo {author} {\bibfnamefont {F.}~\bibnamefont
  {Canet}}, \bibinfo {author} {\bibfnamefont {S.}~\bibnamefont {Mangin}},
  \bibinfo {author} {\bibfnamefont {C.}~\bibnamefont {Bellouard}}, \ and\
  \bibinfo {author} {\bibfnamefont {M.}~\bibnamefont {Piecuch}},\ }\href
  {\doibase 10.1209/epl/i2000-00479-1} {\bibfield  {journal} {\bibinfo
  {journal} {Europhys. Lett.}\ }\textbf {\bibinfo {volume} {52}},\ \bibinfo
  {pages} {594} (\bibinfo {year} {2000})}\BibitemShut {NoStop}%
\bibitem [{\citenamefont {Hebler}\ \emph
  {et~al.}(2016{\natexlab{a}})\citenamefont {Hebler}, \citenamefont
  {B\"ottger}, \citenamefont {Nissen}, \citenamefont {Abrudan}, \citenamefont
  {Radu},\ and\ \citenamefont {Albrecht}}]{PhysRevB.93.184423}%
  \BibitemOpen
  \bibfield  {author} {\bibinfo {author} {\bibfnamefont {B.}~\bibnamefont
  {Hebler}}, \bibinfo {author} {\bibfnamefont {S.}~\bibnamefont {B\"ottger}},
  \bibinfo {author} {\bibfnamefont {D.}~\bibnamefont {Nissen}}, \bibinfo
  {author} {\bibfnamefont {R.}~\bibnamefont {Abrudan}}, \bibinfo {author}
  {\bibfnamefont {F.}~\bibnamefont {Radu}}, \ and\ \bibinfo {author}
  {\bibfnamefont {M.}~\bibnamefont {Albrecht}},\ }\href {\doibase
  10.1103/PhysRevB.93.184423} {\bibfield  {journal} {\bibinfo  {journal} {Phys.
  Rev. B}\ }\textbf {\bibinfo {volume} {93}},\ \bibinfo {pages} {184423}
  (\bibinfo {year} {2016}{\natexlab{a}})}\BibitemShut {NoStop}%
\bibitem [{\citenamefont {Dieny}\ \emph {et~al.}(1991)\citenamefont {Dieny},
  \citenamefont {Speriosu}, \citenamefont {Parkin}, \citenamefont {Gurney},
  \citenamefont {Wilhoit},\ and\ \citenamefont {Mauri}}]{Dieny1991}%
  \BibitemOpen
  \bibfield  {author} {\bibinfo {author} {\bibfnamefont {B.}~\bibnamefont
  {Dieny}}, \bibinfo {author} {\bibfnamefont {V.~S.}\ \bibnamefont {Speriosu}},
  \bibinfo {author} {\bibfnamefont {S.~S.}\ \bibnamefont {Parkin}}, \bibinfo
  {author} {\bibfnamefont {B.~A.}\ \bibnamefont {Gurney}}, \bibinfo {author}
  {\bibfnamefont {D.~R.}\ \bibnamefont {Wilhoit}}, \ and\ \bibinfo {author}
  {\bibfnamefont {D.}~\bibnamefont {Mauri}},\ }\href {\doibase
  10.1103/PhysRevB.43.1297} {\bibfield  {journal} {\bibinfo  {journal} {Phys.
  Rev. B}\ }\textbf {\bibinfo {volume} {43}},\ \bibinfo {pages} {1297}
  (\bibinfo {year} {1991})}\BibitemShut {NoStop}%
\bibitem [{\citenamefont {Radu}\ \emph {et~al.}(2012)\citenamefont {Radu},
  \citenamefont {Abrudan}, \citenamefont {Radu}, \citenamefont {Schmitz},\ and\
  \citenamefont {Zabel}}]{Radu2012}%
  \BibitemOpen
  \bibfield  {author} {\bibinfo {author} {\bibfnamefont {F.}~\bibnamefont
  {Radu}}, \bibinfo {author} {\bibfnamefont {R.}~\bibnamefont {Abrudan}},
  \bibinfo {author} {\bibfnamefont {I.}~\bibnamefont {Radu}}, \bibinfo {author}
  {\bibfnamefont {D.}~\bibnamefont {Schmitz}}, \ and\ \bibinfo {author}
  {\bibfnamefont {H.}~\bibnamefont {Zabel}},\ }\href {\doibase
  10.1038/ncomms1728} {\bibfield  {journal} {\bibinfo  {journal} {Nat.
  Commun.}\ }\textbf {\bibinfo {volume} {3}},\ \bibinfo {pages} {715} (\bibinfo
  {year} {2012})}\BibitemShut {NoStop}%
\bibitem [{\citenamefont {Redon}\ and\ \citenamefont
  {Freitas}(1998)}]{doi:10.1063/1.366995}%
  \BibitemOpen
  \bibfield  {author} {\bibinfo {author} {\bibfnamefont {O.}~\bibnamefont
  {Redon}}\ and\ \bibinfo {author} {\bibfnamefont {P.~P.}\ \bibnamefont
  {Freitas}},\ }\href@noop {} {\bibfield  {journal} {\bibinfo  {journal} {J.
  Appl. Phys.}\ }\textbf {\bibinfo {volume} {83}},\ \bibinfo {pages} {2851}
  (\bibinfo {year} {1998})}\BibitemShut {NoStop}%
\bibitem [{\citenamefont {Lai}\ \emph {et~al.}(2005)\citenamefont {Lai},
  \citenamefont {Wu}, \citenamefont {Lin},\ and\ \citenamefont
  {Huang}}]{doi:10.1063/1.1851954}%
  \BibitemOpen
  \bibfield  {author} {\bibinfo {author} {\bibfnamefont {C.-H.}\ \bibnamefont
  {Lai}}, \bibinfo {author} {\bibfnamefont {Z.-H.}\ \bibnamefont {Wu}},
  \bibinfo {author} {\bibfnamefont {C.-C.}\ \bibnamefont {Lin}}, \ and\
  \bibinfo {author} {\bibfnamefont {P.~H.}\ \bibnamefont {Huang}},\ }\href@noop
  {} {\bibfield  {journal} {\bibinfo  {journal} {J. Appl. Phys.}\ }\textbf
  {\bibinfo {volume} {97}},\ \bibinfo {pages} {10C511} (\bibinfo {year}
  {2005})}\BibitemShut {NoStop}%
\bibitem [{\citenamefont {Liao}\ \emph {et~al.}(2011)\citenamefont {Liao},
  \citenamefont {He}, \citenamefont {Zhang}, \citenamefont {Ma},\ and\
  \citenamefont {Jin}}]{doi:10.1063/1.3536476}%
  \BibitemOpen
  \bibfield  {author} {\bibinfo {author} {\bibfnamefont {J.}~\bibnamefont
  {Liao}}, \bibinfo {author} {\bibfnamefont {H.}~\bibnamefont {He}}, \bibinfo
  {author} {\bibfnamefont {Z.}~\bibnamefont {Zhang}}, \bibinfo {author}
  {\bibfnamefont {B.}~\bibnamefont {Ma}}, \ and\ \bibinfo {author}
  {\bibfnamefont {Q.~Y.}\ \bibnamefont {Jin}},\ }\href@noop {} {\bibfield
  {journal} {\bibinfo  {journal} {J. Appl. Phys.}\ }\textbf {\bibinfo {volume}
  {109}},\ \bibinfo {pages} {023907} (\bibinfo {year} {2011})}\BibitemShut
  {NoStop}%
\bibitem [{\citenamefont {Lin}\ \emph {et~al.}(2006)\citenamefont {Lin},
  \citenamefont {Lai}, \citenamefont {Liao}, \citenamefont {Wu}, \citenamefont
  {Huang},\ and\ \citenamefont {Jiang}}]{doi:10.1063/1.2172193}%
  \BibitemOpen
  \bibfield  {author} {\bibinfo {author} {\bibfnamefont {M.-S.}\ \bibnamefont
  {Lin}}, \bibinfo {author} {\bibfnamefont {C.-H.}\ \bibnamefont {Lai}},
  \bibinfo {author} {\bibfnamefont {Y.-Y.}\ \bibnamefont {Liao}}, \bibinfo
  {author} {\bibfnamefont {Z.-H.}\ \bibnamefont {Wu}}, \bibinfo {author}
  {\bibfnamefont {S.-H.}\ \bibnamefont {Huang}}, \ and\ \bibinfo {author}
  {\bibfnamefont {R.-F.}\ \bibnamefont {Jiang}},\ }\href@noop {} {\bibfield
  {journal} {\bibinfo  {journal} {J. Appl. Phys.}\ }\textbf {\bibinfo {volume}
  {99}},\ \bibinfo {pages} {08T106} (\bibinfo {year} {2006})}\BibitemShut
  {NoStop}%
\bibitem [{\citenamefont {Moodera}\ \emph {et~al.}(1995)\citenamefont
  {Moodera}, \citenamefont {Kinder}, \citenamefont {Wong},\ and\ \citenamefont
  {Meservey}}]{Moodera1995}%
  \BibitemOpen
  \bibfield  {author} {\bibinfo {author} {\bibfnamefont {J.~S.}\ \bibnamefont
  {Moodera}}, \bibinfo {author} {\bibfnamefont {L.~R.}\ \bibnamefont {Kinder}},
  \bibinfo {author} {\bibfnamefont {T.~M.}\ \bibnamefont {Wong}}, \ and\
  \bibinfo {author} {\bibfnamefont {R.}~\bibnamefont {Meservey}},\ }\href
  {\doibase 10.1103/PhysRevLett.74.3273} {\bibfield  {journal} {\bibinfo
  {journal} {Phys. Rev. Lett.}\ }\textbf {\bibinfo {volume} {74}},\ \bibinfo
  {pages} {3273} (\bibinfo {year} {1995})}\BibitemShut {NoStop}%
\bibitem [{\citenamefont {Romer}\ \emph {et~al.}(2012)\citenamefont {Romer},
  \citenamefont {Marioni}, \citenamefont {Thorwarth}, \citenamefont {Joshi},
  \citenamefont {Corticelli}, \citenamefont {Hug}, \citenamefont {Oezer},
  \citenamefont {Parlinska-Wojtan},\ and\ \citenamefont
  {Rohrmann}}]{romer_temperature_2012}%
  \BibitemOpen
  \bibfield  {author} {\bibinfo {author} {\bibfnamefont {S.}~\bibnamefont
  {Romer}}, \bibinfo {author} {\bibfnamefont {M.~A.}\ \bibnamefont {Marioni}},
  \bibinfo {author} {\bibfnamefont {K.}~\bibnamefont {Thorwarth}}, \bibinfo
  {author} {\bibfnamefont {N.~R.}\ \bibnamefont {Joshi}}, \bibinfo {author}
  {\bibfnamefont {C.~E.}\ \bibnamefont {Corticelli}}, \bibinfo {author}
  {\bibfnamefont {H.~J.}\ \bibnamefont {Hug}}, \bibinfo {author} {\bibfnamefont
  {S.}~\bibnamefont {Oezer}}, \bibinfo {author} {\bibfnamefont
  {M.}~\bibnamefont {Parlinska-Wojtan}}, \ and\ \bibinfo {author}
  {\bibfnamefont {H.}~\bibnamefont {Rohrmann}},\ }\href@noop {} {\bibfield
  {journal} {\bibinfo  {journal} {Appl. Phys. Lett.}\ }\textbf {\bibinfo
  {volume} {101}},\ \bibinfo {pages} {222404} (\bibinfo {year}
  {2012})}\BibitemShut {NoStop}%
\bibitem [{\citenamefont {Schubert}\ \emph {et~al.}(2013)\citenamefont
  {Schubert}, \citenamefont {Hebler}, \citenamefont {Schletter}, \citenamefont
  {Liebig}, \citenamefont {Daniel}, \citenamefont {Abrudan}, \citenamefont
  {Radu},\ and\ \citenamefont {Albrecht}}]{schubert_interfacial_2013}%
  \BibitemOpen
  \bibfield  {author} {\bibinfo {author} {\bibfnamefont {C.}~\bibnamefont
  {Schubert}}, \bibinfo {author} {\bibfnamefont {B.}~\bibnamefont {Hebler}},
  \bibinfo {author} {\bibfnamefont {H.}~\bibnamefont {Schletter}}, \bibinfo
  {author} {\bibfnamefont {A.}~\bibnamefont {Liebig}}, \bibinfo {author}
  {\bibfnamefont {M.}~\bibnamefont {Daniel}}, \bibinfo {author} {\bibfnamefont
  {R.}~\bibnamefont {Abrudan}}, \bibinfo {author} {\bibfnamefont
  {F.}~\bibnamefont {Radu}}, \ and\ \bibinfo {author} {\bibfnamefont
  {M.}~\bibnamefont {Albrecht}},\ }\href {\doibase 10.1103/PhysRevB.87.054415}
  {\bibfield  {journal} {\bibinfo  {journal} {Phys. Rev. B}\ }\textbf {\bibinfo
  {volume} {87}},\ \bibinfo {pages} {054415} (\bibinfo {year}
  {2013})}\BibitemShut {NoStop}%
\bibitem [{\citenamefont {Hebler}\ \emph {et~al.}(2017)\citenamefont {Hebler},
  \citenamefont {Reinhardt}, \citenamefont {Katona}, \citenamefont {Hellwig},\
  and\ \citenamefont {Albrecht}}]{Hebler2017}%
  \BibitemOpen
  \bibfield  {author} {\bibinfo {author} {\bibfnamefont {B.}~\bibnamefont
  {Hebler}}, \bibinfo {author} {\bibfnamefont {P.}~\bibnamefont {Reinhardt}},
  \bibinfo {author} {\bibfnamefont {G.~L.}\ \bibnamefont {Katona}}, \bibinfo
  {author} {\bibfnamefont {O.}~\bibnamefont {Hellwig}}, \ and\ \bibinfo
  {author} {\bibfnamefont {M.}~\bibnamefont {Albrecht}},\ }\href {\doibase
  10.1103/PhysRevB.95.104410} {\bibfield  {journal} {\bibinfo  {journal} {Phys.
  Rev. B}\ }\textbf {\bibinfo {volume} {95}},\ \bibinfo {pages} {104410}
  (\bibinfo {year} {2017})}\BibitemShut {NoStop}%
\bibitem [{\citenamefont {Vas'kovskiy}\ \emph {et~al.}(2012)\citenamefont
  {Vas'kovskiy}, \citenamefont {Svalov}, \citenamefont {Balymov},\ and\
  \citenamefont {Kulesh}}]{Vaskovskiy2012}%
  \BibitemOpen
  \bibfield  {author} {\bibinfo {author} {\bibfnamefont {V.~O.}\ \bibnamefont
  {Vas'kovskiy}}, \bibinfo {author} {\bibfnamefont {A.~V.}\ \bibnamefont
  {Svalov}}, \bibinfo {author} {\bibfnamefont {K.~G.}\ \bibnamefont {Balymov}},
  \ and\ \bibinfo {author} {\bibfnamefont {N.~A.}\ \bibnamefont {Kulesh}},\
  }\href {\doibase 10.1134/S0031918X1209013X} {\bibfield  {journal} {\bibinfo
  {journal} {Phys. Metals and Metallography}\ }\textbf {\bibinfo {volume}
  {113}},\ \bibinfo {pages} {862} (\bibinfo {year} {2012})}\BibitemShut
  {NoStop}%
\bibitem [{\citenamefont {Vogler}\ \emph {et~al.}(2020)\citenamefont {Vogler},
  \citenamefont {Heigl}, \citenamefont {Mandru}, \citenamefont {Hebler},
  \citenamefont {Marioni}, \citenamefont {Hug}, \citenamefont {Albrecht},\ and\
  \citenamefont {Suess}}]{vogler2020hysteresisfree}%
  \BibitemOpen
  \bibfield  {author} {\bibinfo {author} {\bibfnamefont {C.}~\bibnamefont
  {Vogler}}, \bibinfo {author} {\bibfnamefont {M.}~\bibnamefont {Heigl}},
  \bibinfo {author} {\bibfnamefont {A.-O.}\ \bibnamefont {Mandru}}, \bibinfo
  {author} {\bibfnamefont {B.}~\bibnamefont {Hebler}}, \bibinfo {author}
  {\bibfnamefont {M.}~\bibnamefont {Marioni}}, \bibinfo {author} {\bibfnamefont
  {H.~J.}\ \bibnamefont {Hug}}, \bibinfo {author} {\bibfnamefont
  {M.}~\bibnamefont {Albrecht}}, \ and\ \bibinfo {author} {\bibfnamefont
  {D.}~\bibnamefont {Suess}},\ }\href@noop {} {\enquote {\bibinfo {title}
  {Hysteresis-free magnetization reversal of exchange-coupled bilayers with
  finite magnetic anisotropy},}\ } (\bibinfo {year} {2020}),\ \Eprint
  {http://arxiv.org/abs/2004.04419} {arXiv:2004.04419 [cond-mat.mtrl-sci]}
  \BibitemShut {NoStop}%
\bibitem [{\citenamefont {Zhao}\ \emph {et~al.}(2019)\citenamefont {Zhao},
  \citenamefont {Mandru}, \citenamefont {Vogler}, \citenamefont {Marioni},
  \citenamefont {Suess},\ and\ \citenamefont {Hug}}]{zhao_magnetization_2019}%
  \BibitemOpen
  \bibfield  {author} {\bibinfo {author} {\bibfnamefont {X.}~\bibnamefont
  {Zhao}}, \bibinfo {author} {\bibfnamefont {A.-O.}\ \bibnamefont {Mandru}},
  \bibinfo {author} {\bibfnamefont {C.}~\bibnamefont {Vogler}}, \bibinfo
  {author} {\bibfnamefont {M.~A.}\ \bibnamefont {Marioni}}, \bibinfo {author}
  {\bibfnamefont {D.}~\bibnamefont {Suess}}, \ and\ \bibinfo {author}
  {\bibfnamefont {H.~J.}\ \bibnamefont {Hug}},\ }\href@noop {} {\bibfield
  {journal} {\bibinfo  {journal} {ACS Appl. Nano Mater.}\ }\textbf {\bibinfo
  {volume} {2}},\ \bibinfo {pages} {7478} (\bibinfo {year} {2019})}\BibitemShut
  {NoStop}%
\bibitem [{\citenamefont {Heigl}\ \emph {et~al.}(2020)\citenamefont {Heigl},
  \citenamefont {Wendler}, \citenamefont {Haugg},\ and\ \citenamefont
  {Albrecht}}]{doi:10.1063/5.0010112}%
  \BibitemOpen
  \bibfield  {author} {\bibinfo {author} {\bibfnamefont {M.}~\bibnamefont
  {Heigl}}, \bibinfo {author} {\bibfnamefont {R.}~\bibnamefont {Wendler}},
  \bibinfo {author} {\bibfnamefont {S.~D.}\ \bibnamefont {Haugg}}, \ and\
  \bibinfo {author} {\bibfnamefont {M.}~\bibnamefont {Albrecht}},\ }\href@noop
  {} {\bibfield  {journal} {\bibinfo  {journal} {J. Appl. Phys.}\ }\textbf
  {\bibinfo {volume} {127}},\ \bibinfo {pages} {233902} (\bibinfo {year}
  {2020})}\BibitemShut {NoStop}%
\bibitem [{\citenamefont {Hug}\ \emph {et~al.}(1999)\citenamefont {Hug},
  \citenamefont {Stiefel}, \citenamefont {van Schendel}, \citenamefont {Moser},
  \citenamefont {Martin},\ and\ \citenamefont
  {Güntherodt}}]{doi:10.1063/1.1149970}%
  \BibitemOpen
  \bibfield  {author} {\bibinfo {author} {\bibfnamefont {H.~J.}\ \bibnamefont
  {Hug}}, \bibinfo {author} {\bibfnamefont {B.}~\bibnamefont {Stiefel}},
  \bibinfo {author} {\bibfnamefont {P.~J.~A.}\ \bibnamefont {van Schendel}},
  \bibinfo {author} {\bibfnamefont {A.}~\bibnamefont {Moser}}, \bibinfo
  {author} {\bibfnamefont {S.}~\bibnamefont {Martin}}, \ and\ \bibinfo {author}
  {\bibfnamefont {H.-J.}\ \bibnamefont {Güntherodt}},\ }\href@noop {}
  {\bibfield  {journal} {\bibinfo  {journal} {Rev. Sci. Instruments}\ }\textbf
  {\bibinfo {volume} {70}},\ \bibinfo {pages} {3625} (\bibinfo {year}
  {1999})}\BibitemShut {NoStop}%
\bibitem [{\citenamefont {Zhao}\ \emph {et~al.}(2018)\citenamefont {Zhao},
  \citenamefont {Schwenk}, \citenamefont {Mandru}, \citenamefont {Penedo},
  \citenamefont {Bacani}, \citenamefont {Marioni},\ and\ \citenamefont
  {Hug}}]{Zhao:2018kd}%
  \BibitemOpen
  \bibfield  {author} {\bibinfo {author} {\bibfnamefont {X.}~\bibnamefont
  {Zhao}}, \bibinfo {author} {\bibfnamefont {J.}~\bibnamefont {Schwenk}},
  \bibinfo {author} {\bibfnamefont {A.~O.}\ \bibnamefont {Mandru}}, \bibinfo
  {author} {\bibfnamefont {M.}~\bibnamefont {Penedo}}, \bibinfo {author}
  {\bibfnamefont {M.}~\bibnamefont {Bacani}}, \bibinfo {author} {\bibfnamefont
  {M.~A.}\ \bibnamefont {Marioni}}, \ and\ \bibinfo {author} {\bibfnamefont
  {H.~J.}\ \bibnamefont {Hug}},\ }\href@noop {} {\bibfield  {journal} {\bibinfo
   {journal} {New J.Phys.}\ }\textbf {\bibinfo {volume} {20}},\ \bibinfo
  {pages} {013018} (\bibinfo {year} {2018})}\BibitemShut {NoStop}%
\bibitem [{\citenamefont {Makarov}\ \emph {et~al.}(2010)\citenamefont
  {Makarov}, \citenamefont {Lee}, \citenamefont {Brombacher}, \citenamefont
  {Schubert}, \citenamefont {Fuger}, \citenamefont {Suess}, \citenamefont
  {Fidler},\ and\ \citenamefont {Albrecht}}]{doi:10.1063/1.3309417}%
  \BibitemOpen
  \bibfield  {author} {\bibinfo {author} {\bibfnamefont {D.}~\bibnamefont
  {Makarov}}, \bibinfo {author} {\bibfnamefont {J.}~\bibnamefont {Lee}},
  \bibinfo {author} {\bibfnamefont {C.}~\bibnamefont {Brombacher}}, \bibinfo
  {author} {\bibfnamefont {C.}~\bibnamefont {Schubert}}, \bibinfo {author}
  {\bibfnamefont {M.}~\bibnamefont {Fuger}}, \bibinfo {author} {\bibfnamefont
  {D.}~\bibnamefont {Suess}}, \bibinfo {author} {\bibfnamefont
  {J.}~\bibnamefont {Fidler}}, \ and\ \bibinfo {author} {\bibfnamefont
  {M.}~\bibnamefont {Albrecht}},\ }\href@noop {} {\bibfield  {journal}
  {\bibinfo  {journal} {Appl. Phys. Lett.}\ }\textbf {\bibinfo {volume} {96}},\
  \bibinfo {pages} {062501} (\bibinfo {year} {2010})}\BibitemShut {NoStop}%
\bibitem [{\citenamefont {Hebler}\ \emph
  {et~al.}(2016{\natexlab{b}})\citenamefont {Hebler}, \citenamefont
  {Hassdenteufel}, \citenamefont {Reinhardt}, \citenamefont {Karl},\ and\
  \citenamefont {Albrecht}}]{10.3389/fmats.2016.00008}%
  \BibitemOpen
  \bibfield  {author} {\bibinfo {author} {\bibfnamefont {B.}~\bibnamefont
  {Hebler}}, \bibinfo {author} {\bibfnamefont {A.}~\bibnamefont
  {Hassdenteufel}}, \bibinfo {author} {\bibfnamefont {P.}~\bibnamefont
  {Reinhardt}}, \bibinfo {author} {\bibfnamefont {H.}~\bibnamefont {Karl}}, \
  and\ \bibinfo {author} {\bibfnamefont {M.}~\bibnamefont {Albrecht}},\ }\href
  {\doibase 10.3389/fmats.2016.00008} {\bibfield  {journal} {\bibinfo
  {journal} {Front. Mater.}\ }\textbf {\bibinfo {volume} {3}},\ \bibinfo
  {pages} {8} (\bibinfo {year} {2016}{\natexlab{b}})}\BibitemShut {NoStop}%
\bibitem [{\citenamefont {Graves}\ \emph {et~al.}(2013)\citenamefont {Graves},
  \citenamefont {Reid}, \citenamefont {Wang}, \citenamefont {Wu}, \citenamefont
  {{De Jong}}, \citenamefont {Vahaplar}, \citenamefont {Radu}, \citenamefont
  {Bernstein}, \citenamefont {Messerschmidt}, \citenamefont {M{\"u}ller},
  \citenamefont {Coffee}, \citenamefont {Bionta}, \citenamefont {Epp},
  \citenamefont {Hartmann}, \citenamefont {Kimmel}, \citenamefont {Hauser},
  \citenamefont {Hartmann}, \citenamefont {Holl}, \citenamefont {Gorke},
  \citenamefont {Mentink}, \citenamefont {Tsukamoto}, \citenamefont {Fognini},
  \citenamefont {Turner}, \citenamefont {Schlotter}, \citenamefont {Rolles},
  \citenamefont {Soltau}, \citenamefont {Str{\"u}der}, \citenamefont
  {Acremann}, \citenamefont {Kimel}, \citenamefont {Kirilyuk}, \citenamefont
  {Rasing}, \citenamefont {St{\"o}hr}, \citenamefont {Scherz},\ and\
  \citenamefont {D{\"u}rr}}]{d1a6f4445d5142d39656b29e436b4366}%
  \BibitemOpen
  \bibfield  {author} {\bibinfo {author} {\bibfnamefont {C.}~\bibnamefont
  {Graves}}, \bibinfo {author} {\bibfnamefont {A.}~\bibnamefont {Reid}},
  \bibinfo {author} {\bibfnamefont {T.}~\bibnamefont {Wang}}, \bibinfo {author}
  {\bibfnamefont {B.}~\bibnamefont {Wu}}, \bibinfo {author} {\bibfnamefont
  {S.}~\bibnamefont {{De Jong}}}, \bibinfo {author} {\bibfnamefont
  {K.}~\bibnamefont {Vahaplar}}, \bibinfo {author} {\bibfnamefont
  {I.}~\bibnamefont {Radu}}, \bibinfo {author} {\bibfnamefont {D.}~\bibnamefont
  {Bernstein}}, \bibinfo {author} {\bibfnamefont {M.}~\bibnamefont
  {Messerschmidt}}, \bibinfo {author} {\bibfnamefont {L.}~\bibnamefont
  {M{\"u}ller}}, \bibinfo {author} {\bibfnamefont {R.}~\bibnamefont {Coffee}},
  \bibinfo {author} {\bibfnamefont {M.}~\bibnamefont {Bionta}}, \bibinfo
  {author} {\bibfnamefont {S.}~\bibnamefont {Epp}}, \bibinfo {author}
  {\bibfnamefont {R.}~\bibnamefont {Hartmann}}, \bibinfo {author}
  {\bibfnamefont {N.}~\bibnamefont {Kimmel}}, \bibinfo {author} {\bibfnamefont
  {G.}~\bibnamefont {Hauser}}, \bibinfo {author} {\bibfnamefont
  {A.}~\bibnamefont {Hartmann}}, \bibinfo {author} {\bibfnamefont
  {P.}~\bibnamefont {Holl}}, \bibinfo {author} {\bibfnamefont {H.}~\bibnamefont
  {Gorke}}, \bibinfo {author} {\bibfnamefont {J.}~\bibnamefont {Mentink}},
  \bibinfo {author} {\bibfnamefont {A.}~\bibnamefont {Tsukamoto}}, \bibinfo
  {author} {\bibfnamefont {A.}~\bibnamefont {Fognini}}, \bibinfo {author}
  {\bibfnamefont {J.}~\bibnamefont {Turner}}, \bibinfo {author} {\bibfnamefont
  {W.}~\bibnamefont {Schlotter}}, \bibinfo {author} {\bibfnamefont
  {D.}~\bibnamefont {Rolles}}, \bibinfo {author} {\bibfnamefont
  {H.}~\bibnamefont {Soltau}}, \bibinfo {author} {\bibfnamefont
  {L.}~\bibnamefont {Str{\"u}der}}, \bibinfo {author} {\bibfnamefont
  {Y.}~\bibnamefont {Acremann}}, \bibinfo {author} {\bibfnamefont
  {A.}~\bibnamefont {Kimel}}, \bibinfo {author} {\bibfnamefont
  {A.}~\bibnamefont {Kirilyuk}}, \bibinfo {author} {\bibfnamefont
  {T.}~\bibnamefont {Rasing}}, \bibinfo {author} {\bibfnamefont
  {J.}~\bibnamefont {St{\"o}hr}}, \bibinfo {author} {\bibfnamefont
  {A.}~\bibnamefont {Scherz}}, \ and\ \bibinfo {author} {\bibfnamefont
  {H.}~\bibnamefont {D{\"u}rr}},\ }\href {\doibase 10.1038/nmat3597} {\bibfield
   {journal} {\bibinfo  {journal} {Nat. Mater.}\ }\textbf {\bibinfo {volume}
  {12}},\ \bibinfo {pages} {293} (\bibinfo {year} {2013})}\BibitemShut
  {NoStop}%
\bibitem [{\citenamefont {Abert}\ \emph {et~al.}(2013)\citenamefont {Abert},
  \citenamefont {Exl}, \citenamefont {Bruckner}, \citenamefont {Drews},\ and\
  \citenamefont {Suess}}]{abert_magnum.fe_2013}%
  \BibitemOpen
  \bibfield  {author} {\bibinfo {author} {\bibfnamefont {C.}~\bibnamefont
  {Abert}}, \bibinfo {author} {\bibfnamefont {L.}~\bibnamefont {Exl}}, \bibinfo
  {author} {\bibfnamefont {F.}~\bibnamefont {Bruckner}}, \bibinfo {author}
  {\bibfnamefont {A.}~\bibnamefont {Drews}}, \ and\ \bibinfo {author}
  {\bibfnamefont {D.}~\bibnamefont {Suess}},\ }\href {\doibase
  10.1016/j.jmmm.2013.05.051} {\bibfield  {journal} {\bibinfo  {journal} {J.
  Magn. Magn. Mater.}\ }\textbf {\bibinfo {volume} {345}},\ \bibinfo {pages}
  {29} (\bibinfo {year} {2013})}\BibitemShut {NoStop}%
\end{thebibliography}%

\end{document}